\documentstyle[12pt,epsf]{article}
\begin{document}
\begin{center}
{\bf Transition to the region of central collisions and  critical phenomena}
\vskip 5mm
M.K. Suleymanov${^1}{^\star}{^\circ} $,O.B. Abdinnov${^1}$,N.S. Angelov${^2}$,A.S. Vodopianov${^2}$.
\vskip 5mm
{\small
(1) {\it
Institute of Physics,Academy of Sciences of Azerbaijan Republic}\\
(2) {\it
Joint Institute for Nuclear Research,
141980, Dubna, Moscow Region, Russia}\\
$\star$ {\it
Now : Laboratory of High Energies,
Joint Institute for Nuclear Research,
141980, Dubna, Moscow Region, Russia}\\
$\circ$ E-mail: mais@sunhe.jinr.ru
}
\end{center}
\vskip 5mm
\begin{center}
\begin{minipage}{150mm}
\centerline{\bf Abstract}

The experimental results  on the relation  between  the processes of total
disintegration  of  nuclei (or  the cases with   central collisions) and the
critical   phenomena which can occur in the region of high degree of nuclear
disintegration or  of collision  centrality are discussed.

\end{minipage}
\end{center}
\vskip 10mm

\section{ Analysis of the events characteristics depending on  disintegration
   degree of nuclei.}

We want to note that to determine the disintegration degree of nuclei or
the centrality degree of collisions the following variables  are usually used
: $n_p$ - number of protons, $n_f$ - number of fragments, $E_{ZD}$ - the
energy flow of secondary particles at an emission angle $\theta =0^0$ and
$E_t$ - the transverse energy of secondary particles.

The research of the processes with total disintegration of nuclei (TDN) was
started long ago in  the experiments with emulsion nuclei [2]. In these
experiments to select the events with TDN  the condition $N_h \ge 28$ is
used. In fig.1  the typical event of the processes of TDN in  $p + Em$ (
reactions at the momentum of  70 GeV/c) is shown. It was found  in 1976. 66
$h$ - particles and 22 $s$ -particles  were identified among the final
reaction products. Particular analysis have demonstrated, that there are 57
one-charge particles ($Z=1$) ,5 two-charge particles ($Z=2$) and 4 particles
with $Z \ge 3$, among  $h$ - particles. Respectively summary charges of $h$ -
particles in this case is the $Z \ge 79$  at insignificant remainder mass. Thus
 the TDN  took place here.

Here we want  to remind that the interest in the processes of TDN was primarily connected with the assumption that high densities of nuclear matter could be realized in these processes and the effects, related to collective properties of nuclear matter, c
ould be observed. However, contrary to
the expectations, one could not receive in the experiment an unequivocal answer to the question on the realization  of these states.

To our opinion, such situation is mainly connected with some methodical difficulties that in the emulsion experiments there was no opportunity to take into account the cases, in which large momentum were transferred to fragments in interactions; the energ
y characteristics of secondary particles were not practically determined, and the statistical material, as a rule, did not exceed some hundred events.

Taking into account all this and also the importance of the problem, the processes with TDN were studied in our experiment [3] according to a new experimental statement. It included the following:

(a) expansion of methodical opportunities of the experiment.  As the bubble chamber technique was used,  we had an opportunity to determine the energy and charge of all secondary particles.

(b) development of new selection criteria  of events with TDN. For this purpose, the idea is used that the processes  with TDN  correspond to qualitatively new states of nuclear matter and the transition to these states occurs in nuclear interactions when
 the number of  protons emitted from nuclei,  Q , reaches a critical value of  $Q^*$. Consequently the presence of $Q^*$ must point to the existence of regime  change in  the behaviour of the characteristics of secondary particles in Q-dependences in the 
region  close to $Q^*$. Hence one can use the following  condition as a selection criterion for the events with  TDN:

$Q \ge Q^*$.

This method of selection of the events with TDN is  experimentally  realized by studying the behaviour of different characteristics of secondary  particles ($a_i$) in hadron-nuclear and nuclear - nuclear interactions depending on $Q$ .

Here  I want to discuss the results of  paper [4] in which the  experimental  data on the $Q$-dependences of  the values of single-particle two-dimensional correlation functions $R$ are presented. The values of $Q$ were determined as

$Q=N_+-N_{\pi^-}$

here $N_+$ and $N_{\pi^-}$ are the numbers of positively charged particles and  $\pi^-$-- mesons , respectively . In such determination $Q$ is a sum charge of an event.  We analyse the

               ${R(x,z)} ={{{d^2N}\over {dxdy}}\over {{{dN}\over {dxdy}}}} - 1$

function. The values of  $R(x,z)$ were normalized in such  way  that the values of  $R=\pm 1$ in the cases of maximum correlation and $R=0$  in the cases when the correlation is absent. Here  +1 corresponds to  positive correlation and -1 to  negative one
.

The experimental data have been obtained from the 2-m propane bubble chamber of  LHE, JINR . In this experiment, we used  5284   $pC$-,  6735   $dC$- , 4852  ${}^4HeC$- and 7327  ${}^{12}CC$- interactions at the momentum of  4.2 A GeV/c ( the methodical d
etails are described in  [8]) . The statistical material was divided into  the groups of events with the following values of  $Q$:

$Q\ge 1 ;2 ; 3 ;...$

The values of $R$ function were determined for $\pi^-$-- mesons and protons.	We have considered the following correlation functions $R(p,\theta), R(p,p_t), R(p,y), R(p,\beta^ 0), R(\theta,p_t)$ ,

\noindent $R(\theta,y), R(\theta,\beta^ 0), R(p_t ,y), R(p_t,\beta^ 0) , R(y,\beta^ 0)$, here :

-  p  is 3-momentum in the laboratory coordinate system (lcs);

- $\theta $  -  emitted angles in the lcs;

- $p_t$ -  transverse momentum;

- y - rapidity in the lcs;

- $\beta^ 0 $ - order of cumulative ( here $\beta^ 0 = {(E-p_L)}/{m_N}$ , E is the total energy (in the lcs),  $p_L$  is the longitudinal momentum (in the lcs) and $m_N$  is the nucleon mass ).

According to the character of the  $Q$ - dependence of  $R$ functions , the data can be divided into two groups : group I - the data on the $Q$ independence of  $R$   and group II - the data showing  the $Q$-dependence of  $R$ . We shall not discuss the r
esults of the first group.

We obtained the following result that  in 90 $\%$ cases  of group II the $Q$-dependence of $R$- function connected with  the variable $p_t$.

The absolute values of $R$- function depending on $Q$ for  group II  are shown  in figures 2-5,  the curves are  hand drawn ). We present the data which point to the regime changes in $Q$-dependences of  $|R|$. The obtained results showd that in 75$\%$ ca
ses , the $Q$-dependence of $R$ has a nonlinear character , i.e. the regime changes take place.  These data allows  to determine the " critical"  values of $Q=Q^*$  corresponding to  the transition from one region to  another.

Thus , we  have that the correlation analysis  confirm that  the events with  TDN qualitatively  differ from "usual" events, and  it could  use the condition

$Q\ge Q^*$

for their separation.

From the figures we can also obtained  the following results:

 	-  in  82 $\%$ of cases for  group II the $R$ function  has the values  $R < 0.3$, i.e.  weak correlations   take place ;

           -  the correlations weaken with the increasing of $A_p$;

           - the character   of the $Q$ - dependence of  $|R|$  also changes  with the increasing  of $A_p$  in the region of  high $Q$  (the region of  TDN). This dependence has  a form of  line with a "break" for $pC$-, $dC$- interactions, it the step-b
y step form for ${}^4HeC$ interactions and a "zigzag" form for  ${}^{12}CC$--interactions.  It is possible that the "zigzag" form is the result of the influence of nuclear fragments, as the  number of the fragments is the  most in the ${}^{12}CC$--interac
tions.

Thus the results, obtained in this paper , have confirmed the assumption that  there is a boundary value 	of $Q^*$ for the quantity  Q the existence of  which  leads to TDN. They also confirmed the idea that  the processes  with TDN  correspond to  qualit
atively new states of nuclear  matter and the transition to these states occurs in nuclear interactions when the number of  protons emitted from the nuclei,  $Q$ , reaches a critical value of  $Q^*$.

\section{ Processes  of total disintegration of nuclei and the cases with  central collisions.}

	Now I want to note that  in  paper [1] it was experimentally shown that the processes with TDN  correspond to the central collisions (CC). The importance of these  investigations is connected with the following. The search for  signals from superdense st
ates of nuclear matter is one of the basic  trends of research in the experiments on relativistic nuclear physics [5]. The best conditions of such states research are the studies of  the events with a maximum number of nucleons - participants in the inter
action  or events connected with CC of nuclei  [6]. To select  such events, the following  criteria are usually used: events with a maximum number   of secondary particles or events  with a minimum flow of secondary  particles energy emitted at  a zero an
gle. Theoretically, both of these conditions must correspond to the value of impact parameter  $b ---> 0$. The "centrality" of nucleus - nucleus collisions is really  a necessary condition for arising the  superdense states of nuclear matter. However, thi
s condition cannot be sufficient as there are processes which are also characterized by a high degree of CC but do not  result in arising superdense states of nuclear matter . In these cases, it is necessary to introduce an additional condition of selecti
on of events to observe a signal from superdense states of nuclear matter. We think that such conditions can be obtained from the research of processes with TDN  [7]. For this purpose, first of all, it is necessary to relate processes with TDN to cases wi
th CC.

	We used the experimental data on the  $pC ,  dC  , {}^4HeC$  and ${}^{12}CC$  interactions at the momentum of 4.2 A GeV/c obtained from the 2-m propane bubble chamber.

  The total statistics of  the events are:

-  8130 events - $pC$ ;

-  6945        -    $dC$ ;

- 11248        -  ${}^4HeC$  and

- 20407        -   ${}^{12}CC$ .

	Methodical details are described in paper [8].

	To relate events with TDN  to the cases with CC, it was supposed that if these events correlated then  with the increasing of the disintegration degree of nuclei -  Q  the average values of the variable K must decrease  and reach a minimum value.  The va
lues of K  are determined in the following way:

\begin{equation}
K = {\sum_{i=1}^{n}{p^2_i} \over \sum_{i=j}^{N}{p^2_j}}
\end{equation}

here $p^2_i$ is the 3-momentum square of an  i-th charged particle with an emission angle $\theta < 5$ and  n  is the number of these particles; $p^2_j$   and  N  are respectively the 3-momentam squared of a  j -th charged particle and the number of all c
harged particles in the event. Q was determined as $Q=N +- N_{\pi^-}$.

	I want to note that  the average values  of impact parameter  $<b>$  must also decrease with the increase of  $Q$   if the condition of  CC, corresponding to the minimum values of  $b$ ,is really achievable. To determine the values of  $<b>$ , we used th
e calculated data on quark-gluon string model[13]

	Fig. 6 shows the $Q$-dependences of the average values  of  $K$.   It is seen that the values of   $<K >$  decreasing with the increase of $Q$ : for  ${}^{12}CC$ interactions in the interval  $Q\ge  6$, for ${}^4HeC--  Q\ge  4$, and for  $dC ,  pC$  - in
teractions in the interval  $Q\ge 3$.

	Thus, one can conclude that in the interval of  high $Q$ , i.e. in the area of TDN,  the average values  of $K$ decreas with the increase of $Q$  and reach their  minimum at the maximum value of $Q$ . From here it follows that the events with TDN corresp
ond to the cases with a minimum flow of energy of charged particles at an emission angle of $\theta < 5^0$.

	One can also see (fig.7) that with the increasing of  $Q$ , the average values  of $b$  decreas  and reaches its minimum at a maximum value of  $Q$ .  This means that the processes with TDN, in the framework of a quark-gluon string model, correspond to e
vents with the highest CC.

\section{Critical phenomenon.}

	To further confirm  the result of the existance of  the points of the regime changes on the behaviour of  some characteristics of secondary particles ai   depending on $Q$ I want to say that at present there are many theoretical papers in which the  proc
esses of nuclear fragmentation [9] and the processes of TDN [10] are considered as a critical phenomenon. Therefore it is possible to  suppose that  if  there exist  cases corresponding to the  critical phenomena among the events with  the TDN then the po
ints of the regime changes in the behaviour  of  some characteristics of secondary particles  depending on the disintegration degree of  collisions - $Q$ could be observed. And I want to add that there are the  experimental data obtained in the proton - n
uclear and nuclear-nuclear interactions at high energy  demonstrate the existence  of the points  of  the regime  changes in the  behaviour of the $a_i =f(Q)$  distributions.

	For example in fig. 8 the average multiplicity of relativistic charged particles depending  on $Q$ is shown for the  ${}^{28}Si_{14} + Em$- reactions at the energies 3.7  and 14.6 GeV per nucleon. To determine the $Q$ a number of charged projectile fragm
ents  $(Z_f)$  were used. The figures were obtained  from  paper [11]. The points of the regime change are observed  in these dependences. These points were used by authors to select the events with CC of nuclei.

	In fig.  9 there are shown the average values of  pseudorapidity $-\eta = -ln tg(\theta /2)$ for $s$-particles(the particles with $\beta > 0.7$)  depending on the number of $g$-particles( the  particles with ($\beta 0.7$ )  for $pEm$-reactions at  the mo
menta of $p_0$=4.5; 24.0;50.0;67.0 and 200.0 GeV/c. This figure was obtained from  paper  [12]. The dashed line in the figure corresponds to the cascade-evaporation model calculations. The points of the regime changes in these distributions are also seen.
 The cascade-evaporation model  calculations cannot describe numerically  these distributions.

	In fig.  10 there are shown the $E_t$-dependences of  the relations of cross sections for the $J/\psi$- production and Drell-Yan processes at the SPS energy (158 GeV per nucleon) for $Pb+Pb$ reactions(the results of  NA50 Collaboration).  The full line c
orresponds to the interactions of light nuclei. The points of the regime changes in these figures are also seen.

	Thus the results clearly demonstrate that  there exit  the points of the regime change of the behavior of $a_i = f(Q)$ distribution which could fix the  region of the  TDN. As I have noted at present there are many papers in which the processes of  nucle
ar fragmentation and the processes of the TDN are considered as the critical phenomena and  for their description a percolation approach was proposed.  We therefore want to use the percolation approaches that as statistical  and percolation theories can d
escribe the critical phenomena best of   all. But the discussed data   showd  that the points of the regime  changes were also observed for interactions of the light nuclei  in which the conditions for  applying of statistical theories are practically  ab
sent.	Supposing these points being connected with  the appearance  of a  critical phenomenon the fragment number change could also have a critical character with the Q increase, because the intermediate nuclear formations (for example percolation cluster)
 could be a source of the nuclear fragmentation.

	To experimentally test this idea we used the experimental data  obtained from the 2-m propane bubble chamber of LHE, JINR. We used 20407 ${}^{12}CC$- interactions at the momentum of 4.2 A GeV/c[8].

	To reach the purpose  we  investigated a number of the events depending on the variable $Q$.  To determine the values of  $Q$ two variants were considered.  In the first variant the values  of $Q$ were determined as

$Q=n_{\pi^+} - n_{\pi^-} + N_p$,

here $n_{\pi^+} , n_{\pi^-}$ and $ N_p$ are the number of identified $\pi^+,\pi^-$ - mesons  and protons respectively . In that determination $Q$ is a number of all the protons in an event without taking into account a remainder of nuclei.  In the second 
 variant the values of $Q$ were determined as

$Q=N_+ - n_{\pi^-}$,

here $N_+$  are  charges  of  all the positively charged particles in an event including  nuclear fragments. In that determination $Q$ is a summary charge of an event.

	 The distributions of the events number depending on $Q$ are shown in fig.11a,b. Here the empty starlets correspond to first variant of $Q$-determination, the full starlets - to second ones (
	 the cases in which the nuclear fragments were  included). There are three regions in these dependences. They are shown by the full lines  drawn  by hand. The first region corresponds to the
	 values  of  $Q\simeq 2$, it is usually named the region of  peripheral collisions.  The second region corresponds to the values  of: $Q\simeq 3- 7$. It is usually named the region of
	 semicentral collisions. The third region corresponds to the values  of $Q\simeq 8- 12$.
It is usually named the region of CC. It is also seen that with the enclusion of fragments number to determine $Q$ the form
of distributions sharply changes and  has a two--steps
structure(full starlets).

	  It is also seen that  the fragments number  to determine the Q
	 being included the form of distributions sharply changes and  has a two-steps structure( full starlets).

	In fig.11b there are shown the Q-dependences of the events  for the calculation data obtained from the quark-gluon string model [13] (QGSM) without nuclear fragments. The empty starlets correspond to the cases in which the stripping protons were not take
n into account and the full starlets correspond to the cases in which the stripping protons were included.  It is seen that the form of the distribution strongly differs from the experimental one in fig.11a. There is no two-steps structure in this figure.
  Therefore we can assert that this difference is  connected with the existence of  nuclear fragments in  ${}^{12}CC$- interactions.

	Thus, the  results demonstrate that the influence of the nuclear fragmentation process in the behaviour of  number of the events depending on $Q$ has a critical character. We suppose this result to be connected with  percolation clusters.

	To confirm the existence of  the percolation cluster  we analysed the angular spectrums - $N_i=  f(cos\theta)$ of identified protons depending on $Q$ and the number of fragments. The used distributions were normalaized on the total number of particles.

	To investigate $Q$-dependences of  $N_i$ -functions we used the following quantities:

${f_1}  =  {{N_4-N_3}\over {N_4+N_3 }}$;

${f_2}  =  {{N_4-N_2}\over { N_4+N_2 }}$;

${f_3} =  {{N_4-N_1}\over { N_4+N_1 }}$.

Here the  $N_1, N_2, N_3$ and $N_4$  are the values of $N_i$  the  groups of the events with the values of  $Q: Q\ge 5$ (this is  peripheral collisions); $Q=6-7; Q=8-9$ and $Q\ge 10$ (this is  CC) respectively. These quantities allow us to compare in deta
il the angular distribution  of protons from the events with  different values of disintegration degree.

	To investigate the values of $f_i$ depending on the number of fragments we also used two variants to determine the variable $Q$.

	Fig.12 shows the $Q$-dependences of  $f_1, f_2$ and $f_3$ as a function of  $cos\theta$. As  well as in fig.11 the empty starlets correspond to the cases in which the nuclear fragments were not taken into account and the full starlets - to the cases in w
hich the nuclear fragments were  included. The difference between two determination  of Q is observed only for the $f_2$ function i.e in the region of the fist step - the region of cluster formation.

	Thus, the  results demonstrate that the influence of  nuclear  fragmentation  process in the behaviour of  the events number  and angular distribution depending on the Q have a critical  character. To explain this result we suppose that it could be conne
cted  with the existence  of percolation clusters.  It is possible to think that with the increase of the centrality degree (fig. 13) the probability of cluster formation grows but further increase the $Q$ (in the region of high $Q$) leads on  the big clu
sters decay  to nuclear fragments and then on  free nucleons. It could be a reason for observation of  two-step structure in the distributions:  the first step connected with the formation of  percolation cluster and the second one - with its decay.

\section{ Summary}

	1. The experimental results obtained in the  proton- nuclear and nuclear-nuclear interactions at high energy  clearly demonstrate the points of  the regime change of the behaviour of some  characteristics of secondary particles depending on the centralit
y degree of collisions. It could  mean that there are  cases corresponding  to the critical  phenomena among the events with  the  central collisions of nuclei. We suppose that   these points could be used to select  the events with the central collisions
 of  nuclei.

	2. For ${}^{12}CC$-interaction  the behaviour of  the number of events, depending on $Q$ also depends on the number of fragments and has a two--steps form. This form is not reproduced by the  calculated data in the framework of the QGS model which does n
ot take into account nuclear fragments. This result as well as the results obtained from the analysis  of angular distributions of protons in peripheral and central collisions could be a confirmation of the existence of percolation clusters.

Finally  I want to say that at GSI, AGS and Nuclotron  energies this result can signal of the existence of the transition of nuclear matter from nucleon states to its  mixed ones. At  SPS, RHIC or LHC energies , a similar result could help to detect "crit
ical" signals of phasetransition  nuclear matter.

	The author consider it his pleasant duty to thank  M.Sumbera and I.Zborovsky for useful discussion and notes , S.Kushpil and V. Kushpil for their help.

\section {References}

 1. M.K. Suleimanov et al.  JINR  Communication, E1-98-328, Dubna,1998.

 2. V.A.  Belyakov et al.  Preprint JINR, P - 331, Dubna, 1959.

 3. Abdinov O.B. et al. JINR  Rapid Communications, N 1[75],1996; Abdinov O.B. et al. JINR Rapid   Communications, N 7[81],1997;Abdinov O.B. et al. JINR Communications, E1-97-342, Dubna, 1997; hep-ex/9712025; Abdinov O.B. et al. JINR Communications, E1-97
-178, Dubna,1997.

4.  M.K. Suleimanov et al., Phys. Rev. C58,351,1998.

5. J.Schukraft and ALICE Collaboration, in: Proceedings X  International Conference on Ultra Relativistic  Nucleus-Nucleus  Collisions, Borlange, Sweden, June 20-24, 1993; Nucl.Phys.,  A566 (1994) 311;   J. W. Harris and STAR Collaboration, in: Proceeding
s X International   Conference on Ultra-Relativistic      Nucleus-Nucleus Collisions, Borlange, Sweden, June  20-24, 1993; Nucl. Phys., A566 (1994) 277.

6. J. Barrette et al. (E814  Collaboration)  Phys. Rev. C50(1994) 3047;G. Wang et al.  (E900  Collaboration)       Phys. Rev. C53(1996) 1811;Y. Akiba et. al. (E802  Collaboration) Phys.  Rev. C56(1997) 1544;B. Hong  et   al.     (FOPI  Collaboration) Phys
. Rev. C57 (1998)  244;Miskowiec et. al (KaoS  Collaboration)  Phys. Rev.   Lett. 72 (1994) 3650;  W.C. Hsi et  al. (ALA DIN Collaboration) Phys. Rev. Lett. 73 (1994)3367;  M.M. Aggarwal  et al.  (WA80,WA93,WA98 Collaboration) Phys. Rev. C56 (1997)  1160;
  T. Alber et al. (NA35  and NA49 Collaborations) Nucl. Phys. A590 (1995) 453;   J. Bachler et.al  (NA35  Collaborations) Z. Phys.      C58 (1993) 541; H. Appelshauser et al.    (NA49     Collaborations) Eur.Phys.J.C2 (1998) 661; Chkhaidze et al.     Phys
ics  Letters B,    volume   411, N 1,2, p.26-32,1997.; L. Ahle, Y. Akiba, et al. Phys. Rev. C, volume 55,      Number 5,      p.2604-2614.1997.

7. K.D. Tolstov, R.A.  Khoshmukhametov. Preprint JINR, P1-6897, Dubna, 1973; Akhrorov et al. Preprint   JINR    P1-9963, Dubna, 1976.; B.Jakobsson,  J.Otterlund,    K.  Kristiansson. Preprint     LUIP-CR-74-14,       Lund, 1974; AA-B-G-D-D-E-K-K-M-P-SP-S-
T-T-UB-U-    Collaboration. Sov.Journal Nucl. Ph.F, v. 55, 4,     1992, p.1010-1020;  Bagdanov V. G. et al. Sov.Journal   Nucl.Ph. v .38, 1983 , p. 1493; Yu. F.Gagarin et al.     Sov.Journal "  News of USSR Academy of Sciences " Phys. Series, v. 38, N 5,1
974, p. 989- 992 ; N.Angelov et al., Sov. Journal Nucl.Ph. v.28, 3 (9), 1978.;  Bondarenko A.I. et al. Sov. Journal Nucl.   Ph. v.60, 11,      1997.; A.  Dabrowska et al.  Phys.  Rev. D47 (1993) 1751.

8. N.Akhababian et al.- JINR Preprint 1-12114, Dubna, 1979.; N.S.Angelov et al.- JINR       Preprint 1-12424,   Dubna, 1989 ; A.I.Bondorenko  et al., JINR Communication, P1-98-292,       Dubna, (1998)

9. J. Desbois, Nucl. Phys. A466, 724 (1987) : J. Nemeth et al. Z.Phys.A 325, 347 (1986); S.  Leray et al.  Nucl.  Phys. A511 (1990) p. 414- 428; A.J. Santiago and K.C. Chung   J.   Phys.G: Nucl. Part. Phys. 16 (1990) p.        1483 -- 1492.

10. X.Campi, J. Desbois  Proc. 23 Int. Winter Meeting on Nucl.Phys; Bormio ,1985;     Bauer W. et al. Nucl.Phys. 1986 .v.452. p.699;A.S. Botvina, L.V. Lanin. Sov. J. Nucl. Phys. 55: 381 - 387, 1992.

11. M.I. Tretyakova. EMU-01 Collaboration. Proceeding of the Xith International Seminar on       High Energy         Physics Problems. Dubna, JINR, 1994.,p.616-626.

12.  S.Vokal, M.Sumbera, JINR Preprint, 1-83-389, Dubna, (1983).

13. N.S. Amelin, L.V.Bravina, Sov. J. Nucl. Phys. 51,211,1990; N.S. Amelin et al.,  Sov. J. Nucl.  Phys.50,272,1990

\section{ Caption of figuers}

{\noindent

Fig.1

Fig.2  Q-dependence of  (R((,pt)(for (-- mesons in pC  , dC  , 4HeC and 12CC
interactions. The values of (R( shown at Q=1;2;3;... correspond to the groups
of events with Q(1;2;3;... , respectively.

Fig.3 Q-dependence of  (R(pt,y)(for (-- mesons in pC  , dC  , 4HeC and 12CC
interactions. The values of (R( shown at Q=1;2;3;... correspond to the groups
of events with Q(1;2;3;... , respectively.

Fig.4 Q-dependence of  (R(p,pt)(for protons in pC  , dC  , 4HeC and 12CC
interactions. The values of (R( shown at Q=1;2;3;... correspond to the groups
of events with Q(1;2;3;... , respectively.

Fig.5 Q-dependence of  (R(pt,y)(for protons in pC  , dC  , 4HeC and 12CC
interactions. The values of (R( shown at Q=1;2;3;... correspond to the groups
of events with Q(1;2;3;... , respectively.

Fig.6 Q  - dependence of the values of  < K > for 12CC ,4HeC , dC  and  pC  interactions.

Fig.7 Q  - dependence of the values of   < b > for 12CC ,4HeC , dC  and  pC  interactions.

Fig.8

Fig.9

Fig.10

Fig.11a,b Q-dependence of the number of events.

Fig.12   Q-dependence of the angular distribution of protons.

Fig.13}

\newpage
\begin{minipage}{4cm}

\end{minipage}

\vskip 8cm
\hspace*{-0cm}
\begin{center}
\hspace*{-0.cm}
\parbox{15cm}{\epsfxsize=15.cm \epsfysize=15.cm\epsfbox[5 5 500 500]
{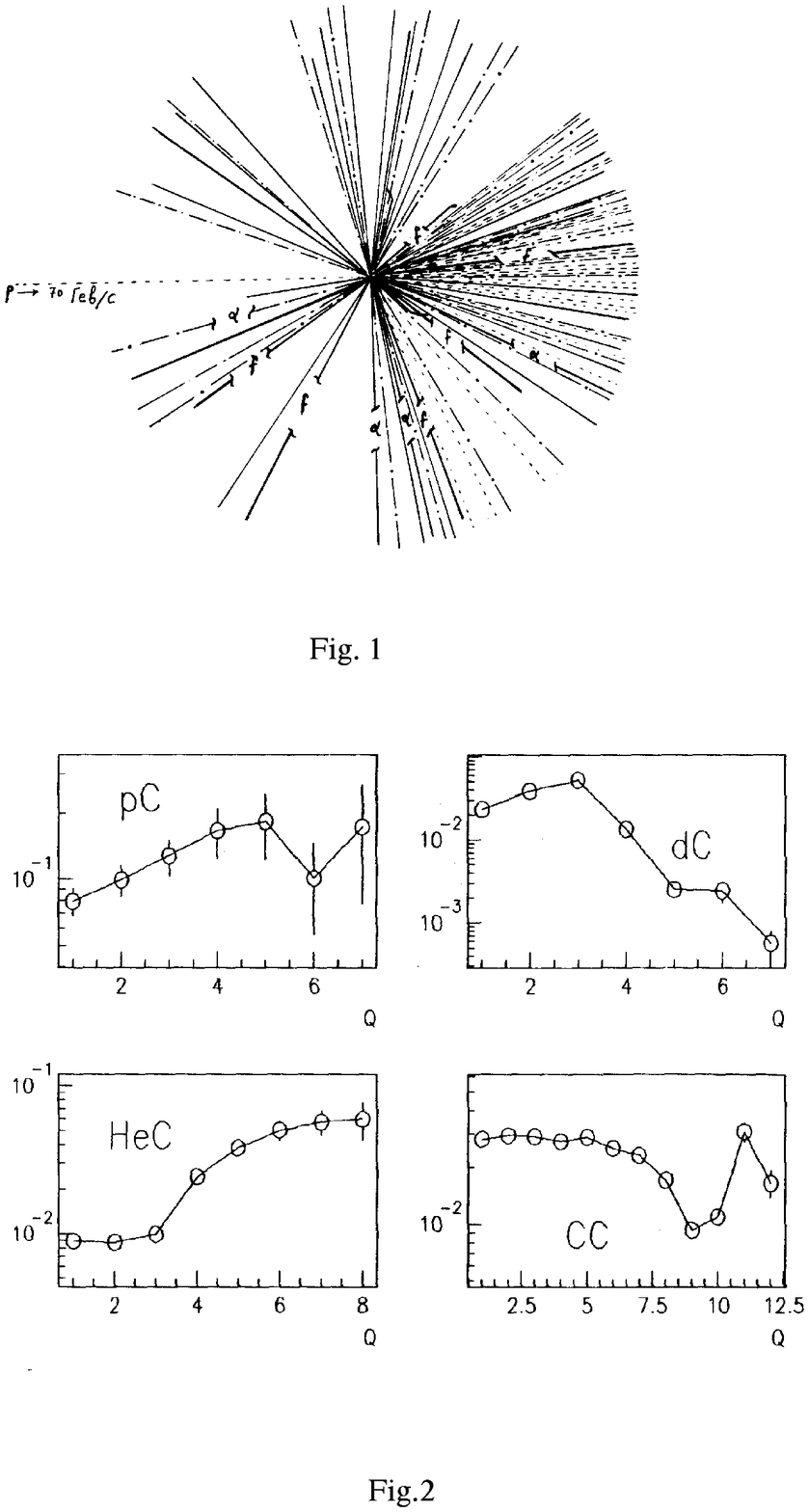}{}}

\vskip 0.25cm

\end{center}
\newpage
\begin{minipage}{4cm}

\end{minipage}

\vskip 8cm
\hspace*{-0cm}
\begin{center}
\hspace*{-0.cm}
\parbox{15cm}{\epsfxsize=15.cm \epsfysize=15.cm\epsfbox[5 5 500 500]
{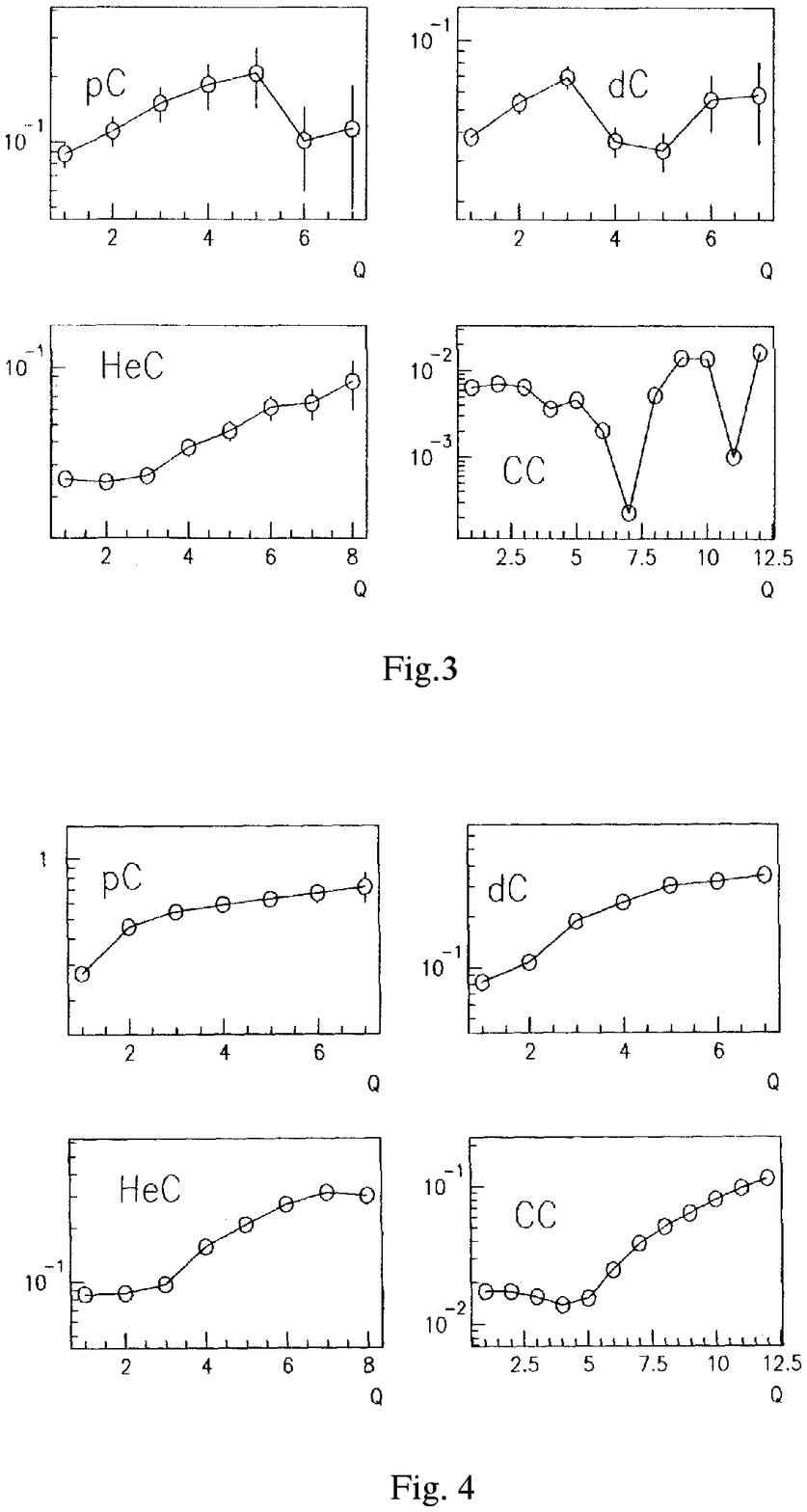}{}}

\vskip 0.25cm

\end{center}
\newpage
\begin{minipage}{4cm}

\end{minipage}

\vskip 8cm
\hspace*{-0cm}
\begin{center}
\hspace*{-0.cm}
\parbox{15cm}{\epsfxsize=15.cm \epsfysize=15.cm\epsfbox[5 5 500 500]
{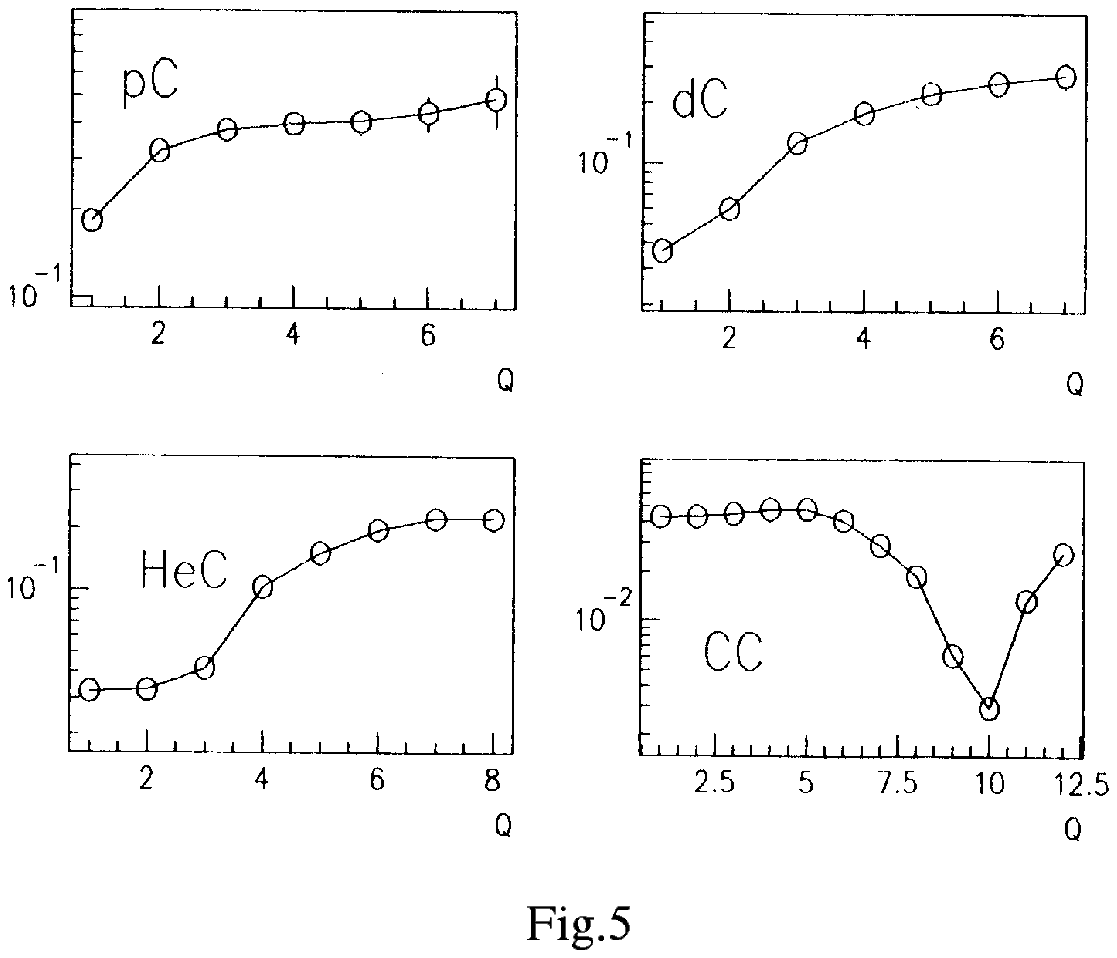}{}}

\vskip 0.25cm

\end{center}
\newpage
\begin{minipage}{4cm}

\end{minipage}

\vskip 8cm
\hspace*{-0cm}
\begin{center}
\hspace*{-0.cm}
\parbox{15cm}{\epsfxsize=15.cm \epsfysize=15.cm\epsfbox[5 5 500 500]
{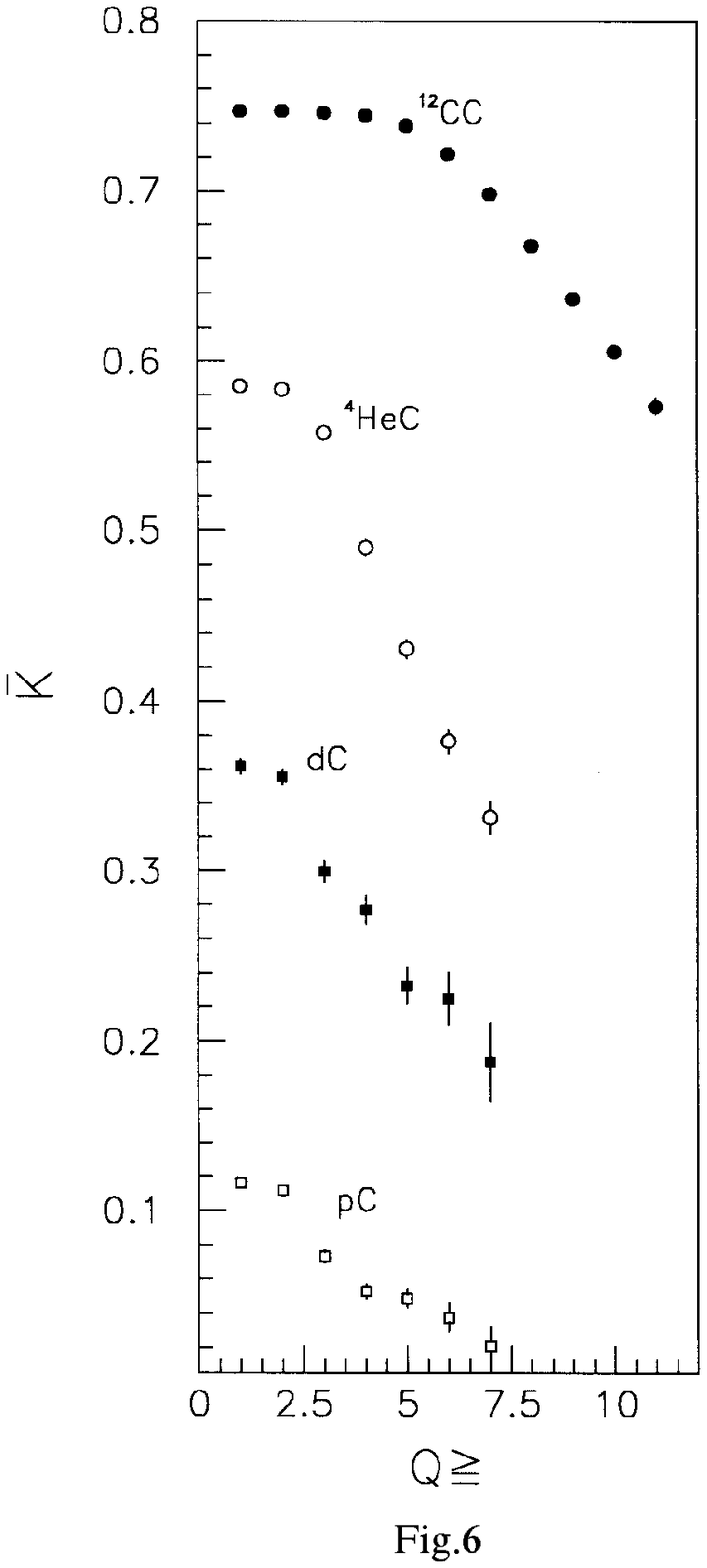}{}}

\vskip 0.25cm

\end{center}
\newpage
\begin{minipage}{4cm}

\end{minipage}

\vskip 8cm
\hspace*{-0cm}
\begin{center}
\hspace*{-0.cm}
\parbox{15cm}{\epsfxsize=15.cm \epsfysize=15.cm\epsfbox[5 5 500 500]
{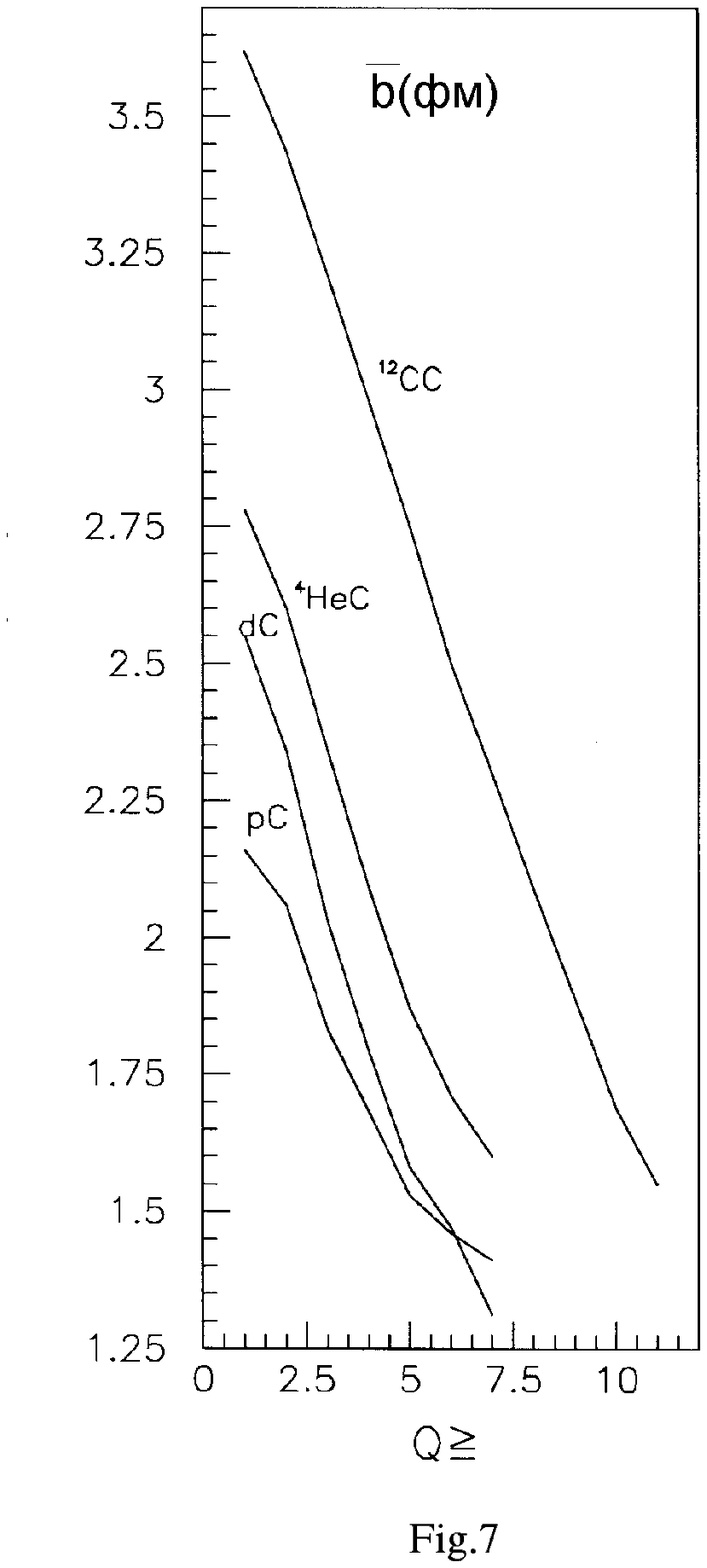}{}}

\vskip 0.25cm

\end{center}

\newpage
\begin{minipage}{4cm}

\end{minipage}

\vskip 4cm
\hspace*{-0cm}
\begin{center}
\hspace*{-0.cm}
\parbox{16cm}{\epsfxsize=16.cm \epsfysize=16.cm\epsfbox[5 5 500 500]
{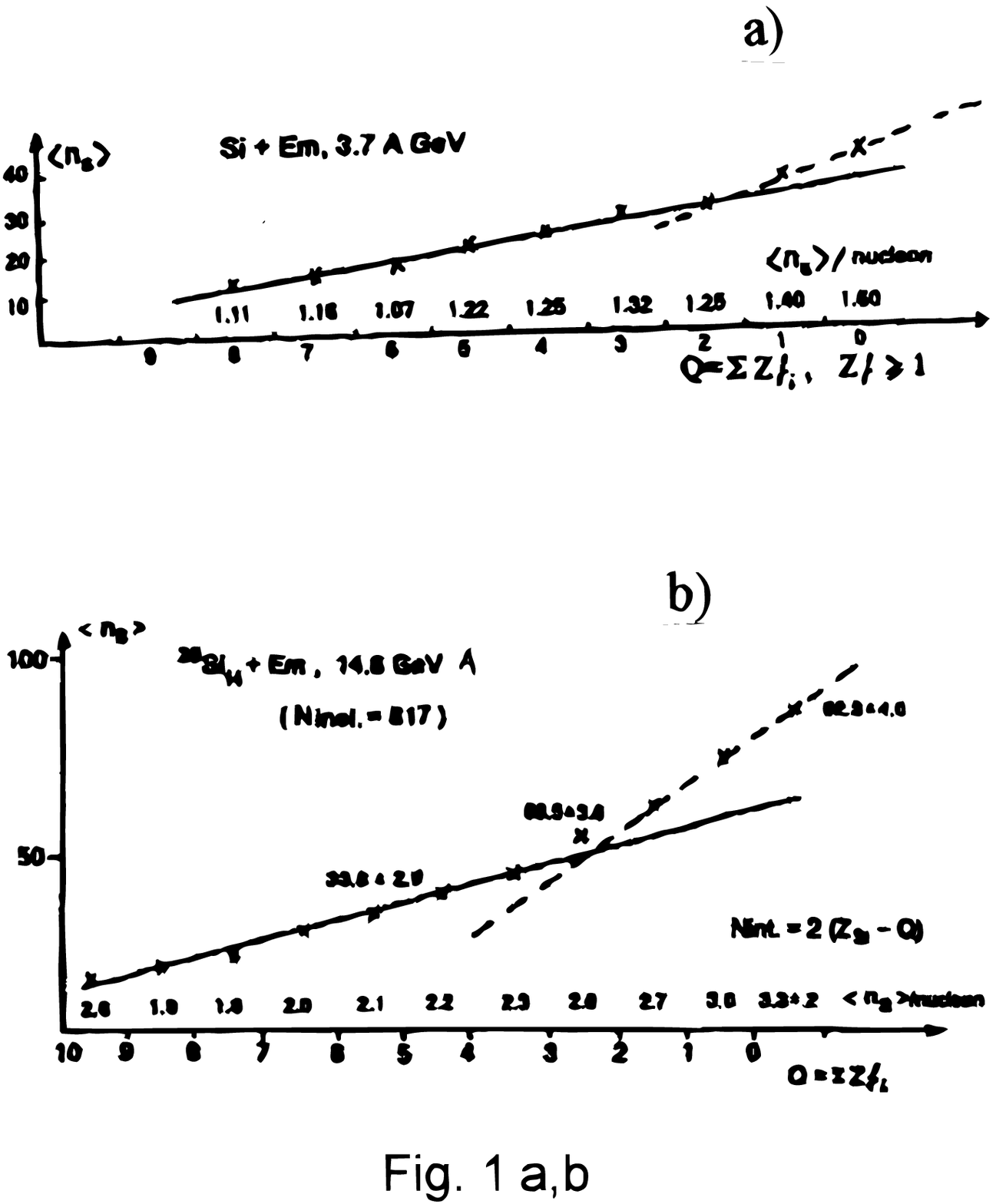}{}}

{\huge \bf Fig.8a,b}

\vskip 0.25cm

\end{center}
\newpage
\begin{minipage}{4cm}

\end{minipage}

\vskip 4cm
\hspace*{-0cm}
\begin{center}
\hspace*{-0.cm}
\parbox{16cm}{\epsfxsize=16.cm \epsfysize=16.cm\epsfbox[5 5 500 500]
{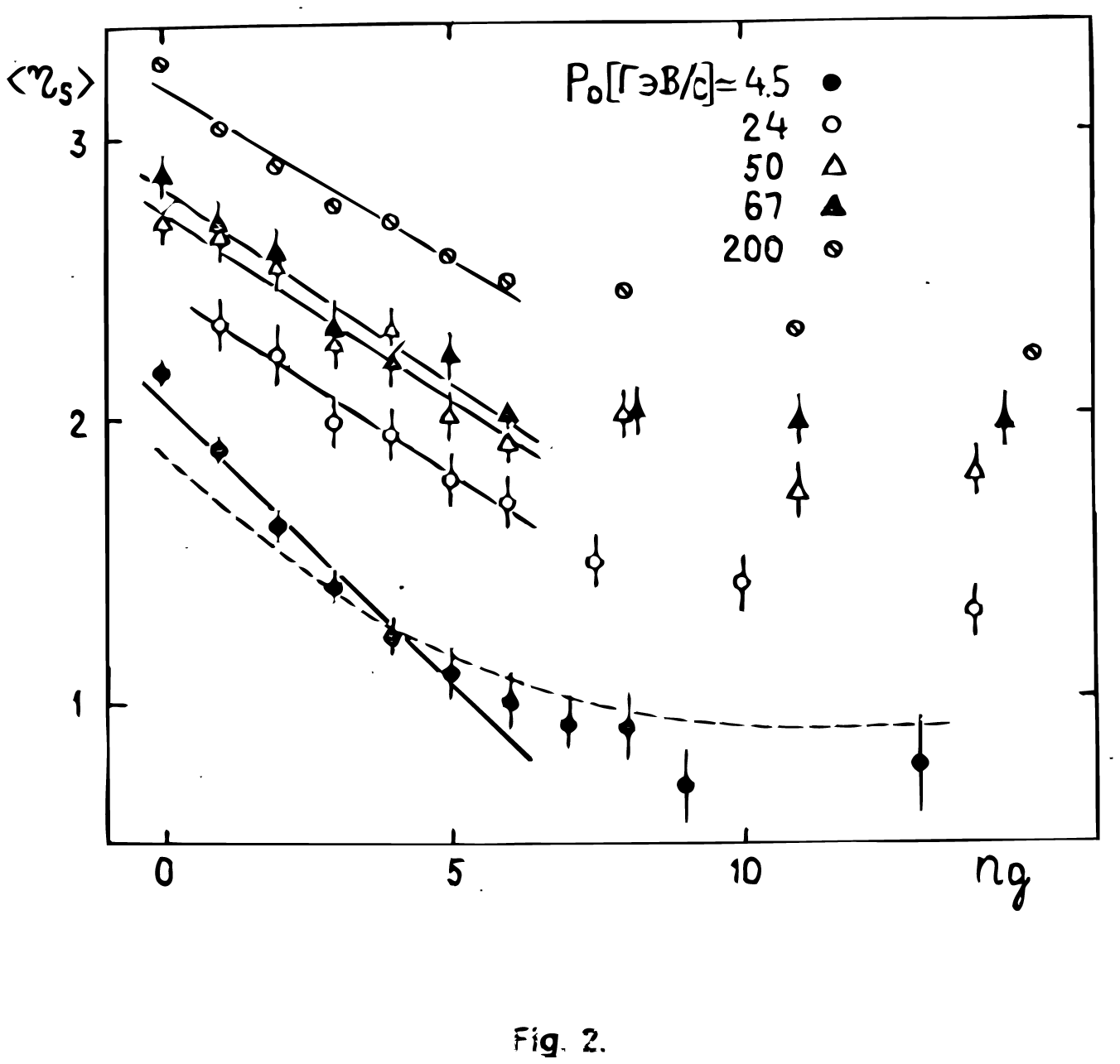}{}}

{\huge \bf Fig.9}

\vskip 0.25cm

\end{center}

\newpage
\begin{minipage}{4cm}

\end{minipage}

\vskip 10cm
\hspace*{-0cm}
\begin{center}
\hspace*{-0.cm}
\parbox{12cm}{\epsfxsize=12.cm \epsfysize=12.cm\epsfbox[5 5 500 500]
{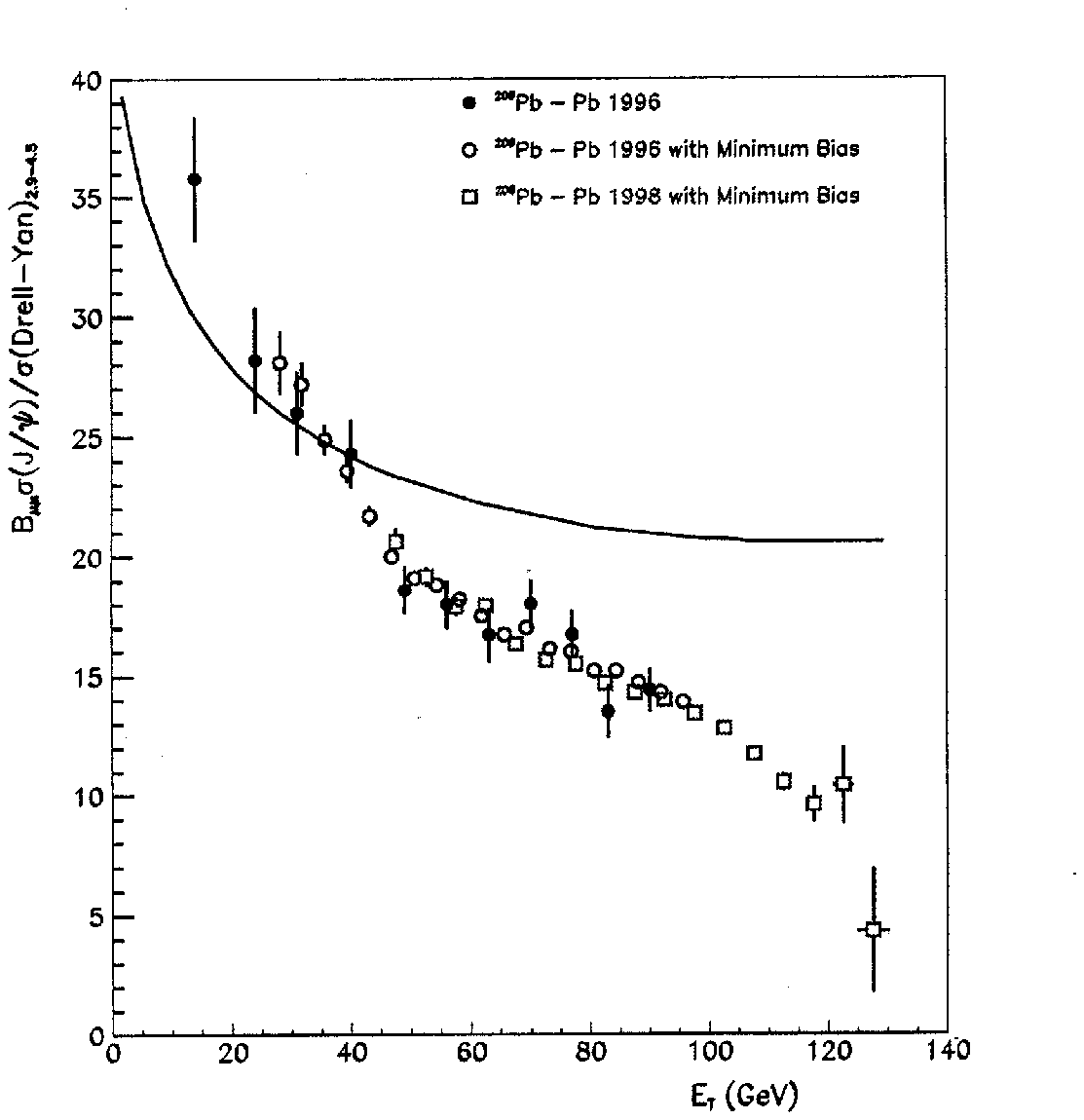}{}}

{\huge \bf Fig.12}

\vskip 0.25cm

\end{center}

\newpage
\begin{minipage}{4cm}

\end{minipage}

\vskip 8cm
\hspace*{-0cm}
\begin{center}
\hspace*{-0.cm}
\parbox{14cm}{\epsfxsize=14.cm \epsfysize=14.cm\epsfbox[5 5 500 500]
{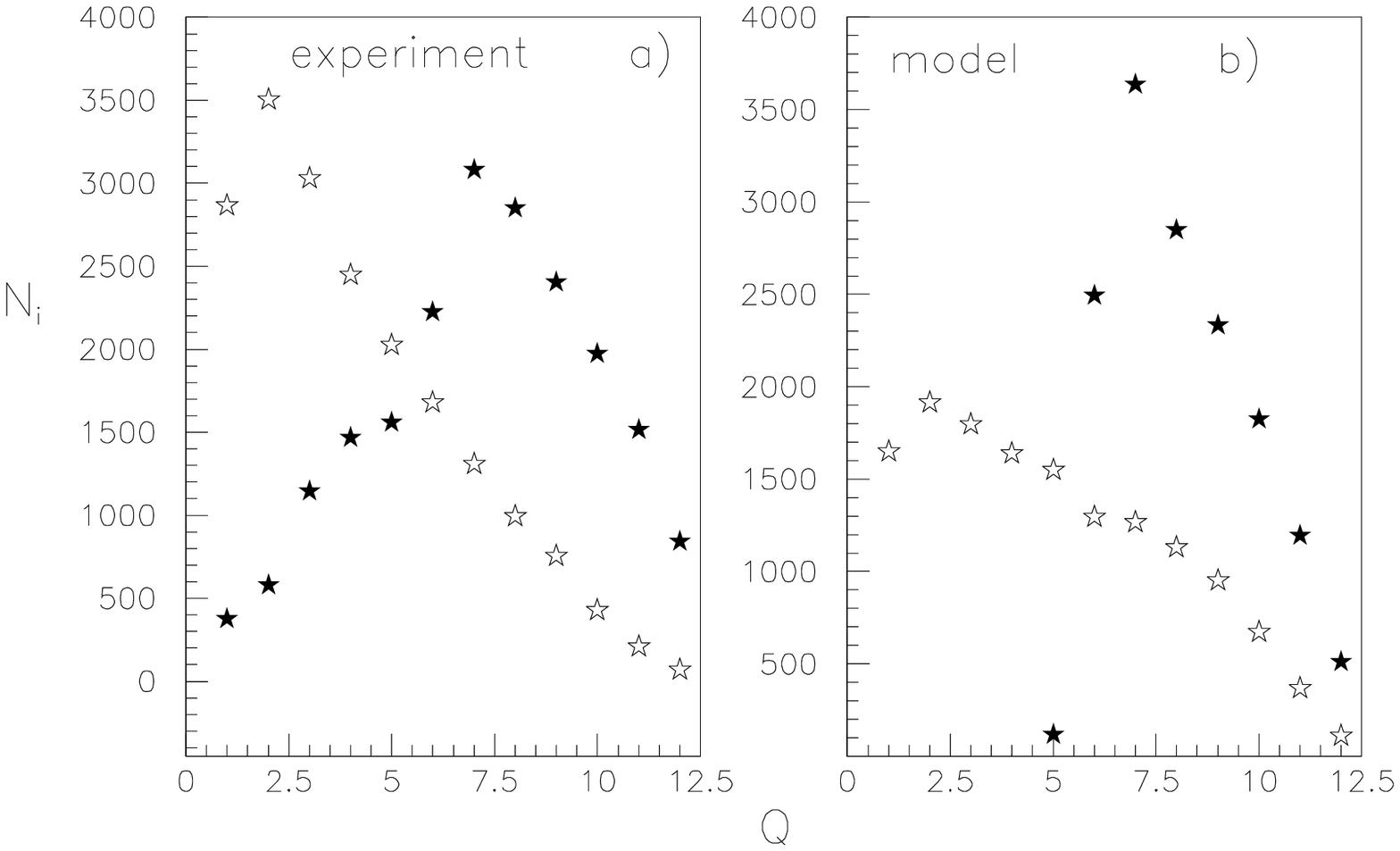}{}}

{\huge \bf Fig.11}

\vskip 0.25cm

\end{center}

\newpage
\begin{minipage}{4cm}

\end{minipage}

\vskip 8cm
\hspace*{-0cm}
\begin{center}
\hspace*{-0.cm}
\parbox{14cm}{\epsfxsize=14.cm \epsfysize=14.cm\epsfbox[5 5 500 500]
{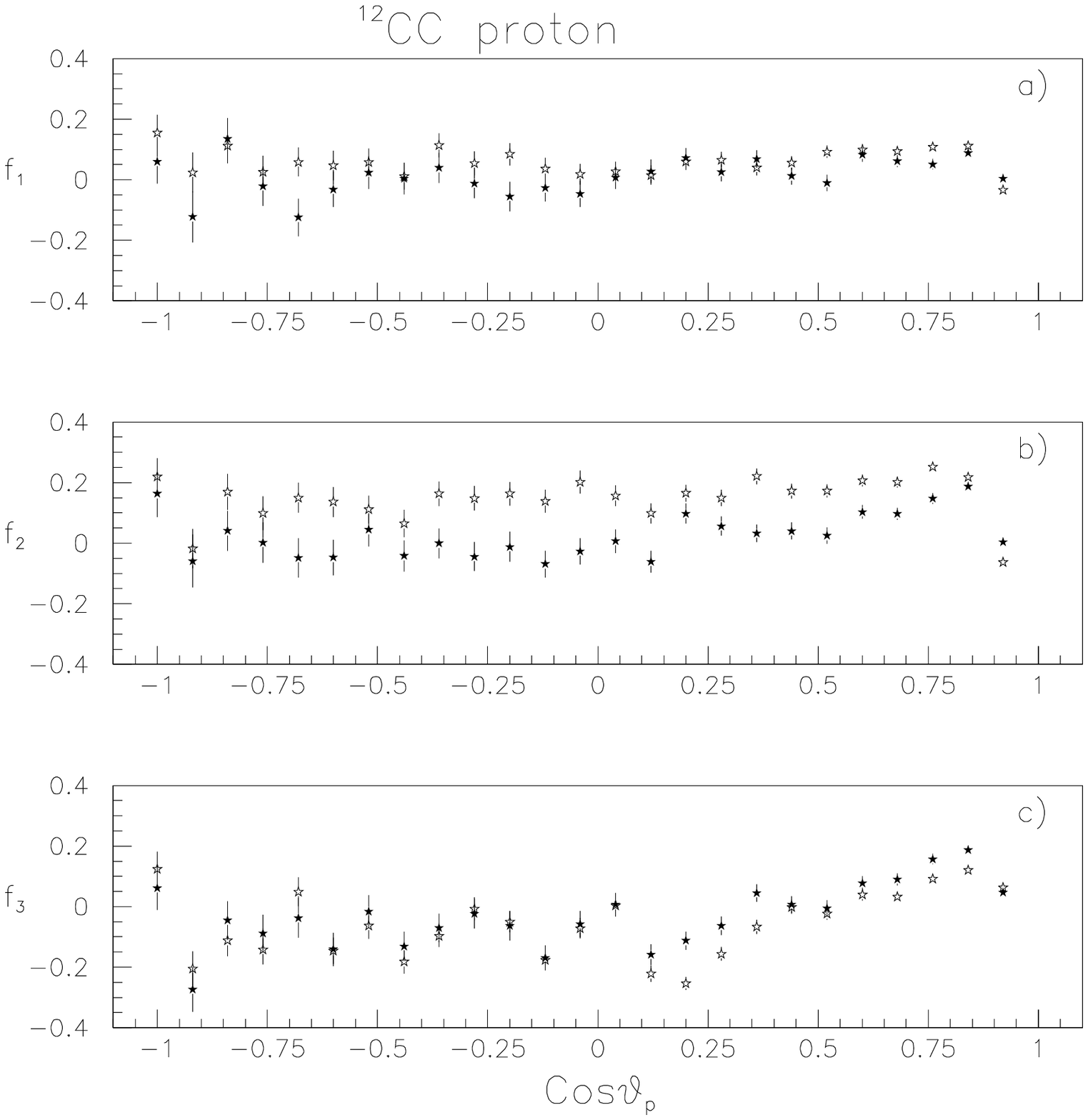}{}}

{\huge \bf Fig.12}

\vskip 0.25cm

\end{center}

\newpage
\begin{minipage}{4cm}

\end{minipage}

\vskip 2cm
\hspace*{-0cm}
\begin{center}
\hspace*{-0.cm}
\parbox{15cm}{\epsfxsize=15.cm \epsfysize=15.cm\epsfbox[5 5 500 500]
{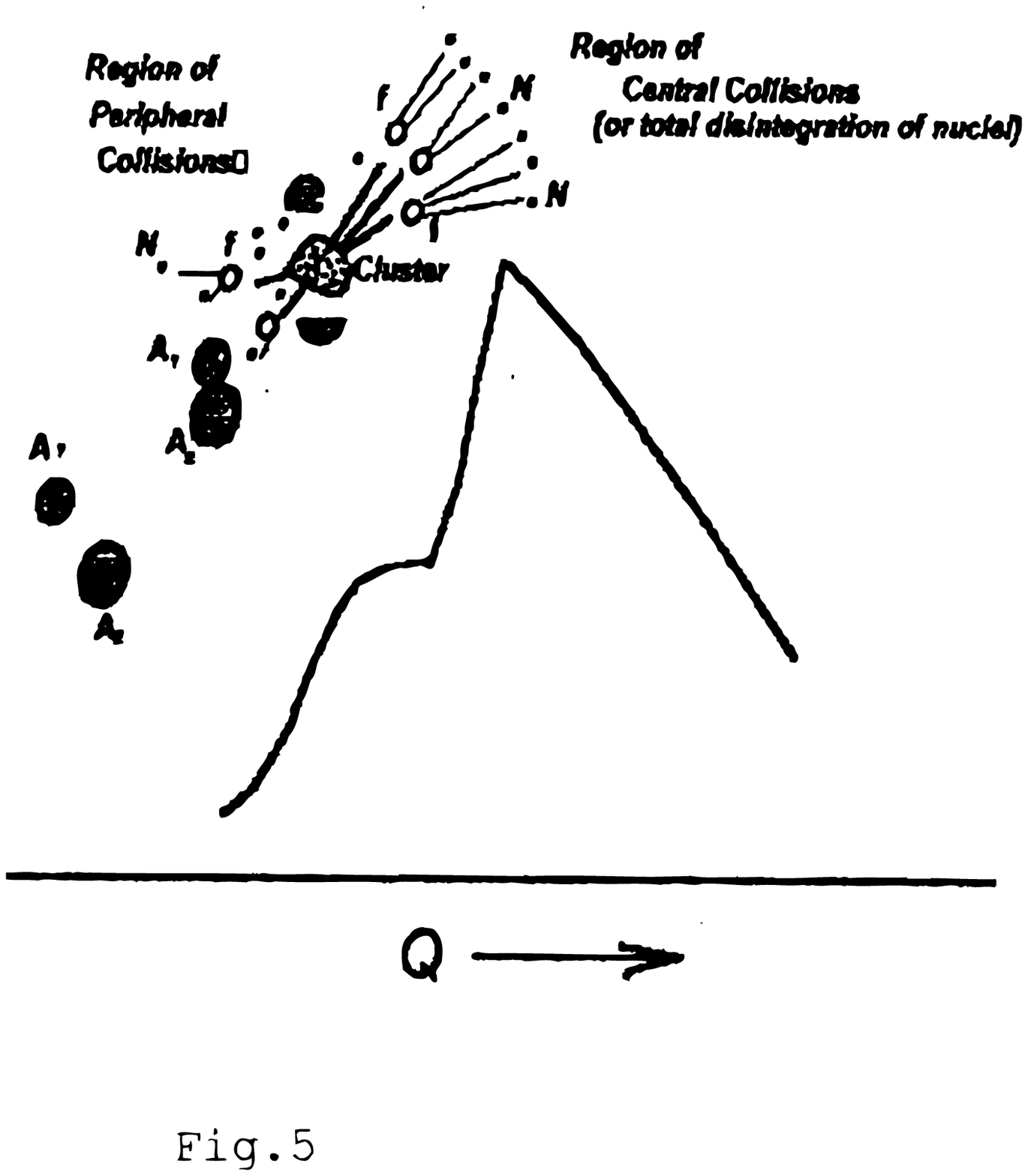}{}}

{\huge \bf fig.13}

\vskip 0.25cm

\end{center}

\end{document}